\pdfoutput=1
\documentclass[prx,twocolumn,showpacs,amsmath,amssymb,floatfix,superscriptaddress]{revtex4-2}

\usepackage{graphicx}  
\usepackage{dcolumn}   
\usepackage{bm}        
\usepackage{amssymb}   
\usepackage{amsmath}
\usepackage{mathtools}
\usepackage{braket}
\usepackage[colorlinks=true, linkcolor=blue, urlcolor=blue,
citecolor=blue]{hyperref}
\newcommand{\uf}{\textcolor{red}{\uparrow}}
\newcommand{\df}{\textcolor{red}{\downarrow}}
\newcommand{\dn}{{\downarrow}}
\newcommand{\up}{{\uparrow}}

\begin{document}
	\title{Electron-mediated  entanglement of two distant  macroscopic ferromagnets within a nonequilibrium spintronic device}
	
	
	\author{A. Suresh}
	\affiliation{Department of Physics and Astronomy, University of Delaware, Newark, DE 19716, USA}
	
	\author{R. D. Soares}
	\affiliation{Departamento de Física e Astronomia, Faculdade de Ciências da Universidade do Porto, Rua do Campo Alegre, s/n, 4169-007 Porto, Portugal}
	
	\author{P. Mondal}
	\affiliation{Department of Physics and Astronomy, University of Delaware, Newark, DE 19716, USA}
	
	\author{J.~P.~Santos~Pires}
	\affiliation{Departamento de Física e Astronomia, Faculdade de Ciências da Universidade do Porto, Rua do Campo Alegre, s/n, 4169-007 Porto, Portugal}
	\affiliation{Centro de Física das Universidades do Minho e do Porto (CF-UM-UP) and Laboratório de Física para Materiais e Tecnologias Emergentes LaPMET, University of Porto, 4169-007 Porto, Portugal}
	
	\author{J.~M.~Viana~Parente~Lopes}
	\affiliation{Departamento de Física e Astronomia, Faculdade de Ciências da Universidade do Porto, Rua do Campo Alegre, s/n, 4169-007 Porto, Portugal}
	\affiliation{Centro de Física das Universidades do Minho e do Porto (CF-UM-UP) and Laboratório de Física para Materiais e Tecnologias Emergentes LaPMET, University of Porto, 4169-007 Porto, Portugal}
	
	\author{Aires Ferreira}
	\affiliation{School of Physics, Engineering and Technology and York Centre for Quantum Technologies, University of York, York YO105DD, United Kingdom}
	
	\author{A. E.~Feiguin}
	\affiliation{Department of Physics, Northeastern University, Boston, MA 02115, USA}
	
	\author{P. Plech\'a\v{c}}
	\affiliation{Department of Mathematical Sciences, University of Delaware, Newark, DE 19716, USA}       
	
	\author{B. K. Nikoli\'c}
	\email{bnikolic@udel.edu}
	\affiliation{Department of Physics and Astronomy, University of Delaware, Newark, DE 19716, USA} 
	
	\begin{abstract}
		Using the nascent concept of quantum spin-transfer torque [A.\,Zholud\,{\em et\,al.},\,Phys.\,Rev.\,Lett.\,{\bf 119}, 257201\,(2017); M.\,D.\,Petrovi\'{c}\,{\em et\,al.},\,Phys.\,Rev.\,X\,{\bf 11},\,021062\,(2021)],\,\,we demonstrate that a current pulse can be harnessed to entangle quantum localized spins of two spatially separated ferromagnets (FMs) which are initially unentangled. The envisaged setup comprises a spin-polarizer (FM$_p$) and   a spin-analyzer (FM$_a$) FM layers separated by normal metal (NM) spacer. The injection of a current pulse into the  device leads to a time-dependent superposition of many-body states characterized by a high degree of entanglement between the spin degrees of freedom of the two distant FM layers. The non-equilibrium dynamics are due to the transfer of spin angular momentum from itinerant electrons to the localized spins via a quantum spin-torque mechanism that remains active even for {\em collinear but antiparallel} arrangements of the FM$_p$ and FM$_a$ magnetizations (a situation in which the conventional spin-torque is absent). We  quantify the mixed-state entanglement generated between the FM layers by tracking the time-evolution of the full density matrix and analyzing the build-up of the mutual logarithmic negativity over time. The effect of decoherence and dissipation in the FM layers due to coupling to bosonic baths at finite temperature, the use of multi-electron current pulses and the dependence on the number of spins are also considered in an effort to ascertain the robustness of our predictions under realistic conditions. Finally, we propose a ``current-pump/X-ray-probe'' scheme, utilizing ultrafast X-ray spectroscopy, that can witness nonequilibrium and transient entanglement of the FM layers by extracting its time-dependent quantum Fisher information.
	\end{abstract}
	\maketitle
	
	\section{Introduction}\label{sec:intro}
	
	Entanglement describes genuinely quantum and nonlocal correlations between different parts of a physical system. Formally, it stems from a many-body wavefunction that is not expressible in a separable fashion, \textit{i.e.}, as the direct product of multiple single-particle states in some basis.  Initially explored in \textit{gedanken} experiments and tests of Bell-type inequalities involving spin-1/2 particles \cite{Einstein1935,Bohm1957,Kiess1993,Kwiat1995},  quantum entanglement nowadays has risen to the forefront of many applications, including quantum cryptography and quantum computation~\cite{Ekert1991, Shor2000, Gisin2002,Bennett98,Stean1998,Beneti2019}. 
	
	The question  of whether quantum entanglement can survive beyond the microscopic domain into the realm of macroscopic phenomena ~\cite{Vedral2004,Morimae2005,Sperling2017} has fascinated physicists since the inception of quantum theory. Even though the laws of quantum physics  are believed to govern the behavior of small quantum units   and large objects alike, the fast decoherence of massive quantum superpositions  makes deviations from a classical description very challenging to detect  on a macroscopic scale \cite{Zurek2003,Joos2003}. Notwithstanding these practical difficulties, continuous experimental efforts over the past two decades in quantum state preparation and readout  of   mechanical systems   have highlighted the possibility to entangle the internal degrees of freedom of larger and larger systems. This includes putting the phonon modes of two distant macroscopic mechanical oscillators (each containing $\gtrsim 10^{12}$ atoms)  into a nonclassical state   ~\cite{Julsgaard2001,Ockeloen2018,Riedinger2018,Kotler2021,Lepinay2021}, even at room temperature~\cite{Lee2011}. Recent experiments have also achieved macroscopic entanglement of a mechanical oscillator (several millimeters long and $\sim 10$ nm thick) with a cloud made up of a billion cesium atoms (a collective atomic spin oscillator) placed at a distance of $\sim 1$ m using photons propagating between the two objects, as an entanglement mediator \cite{Thomas2021}. Besides its fundamental interest~\cite{Leggett2002}, these advances can pave the way to a new class of quantum information technologies and quantum sensors. The possibility to engineer entangled states of large objects also opens up opportunities to improve the sensitive of gravitational wave detectors, such as the  Laser Interferometer Gravitational-Wave Observatory (LIGO), as illustrated by a recent proposal to exploit entanglement between optical fields and atomic clouds to surpass standard quantum limit~\cite{Khalili2018}.

	\begin{figure}
		\includegraphics[width=\linewidth]{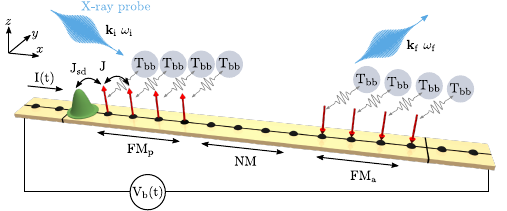}
		\caption{Schematic view of 1D model of a FM$_p$/NM/FM$_a$ spin valve where\,a spin-unpolarized current pulse $I(t)$---carrying charge \mbox{$Q=\int \!\! I(t) dt=N_e e$} comprised of one ($N_e=1$) or more ($N_e>1$) electrons---is injected into the polarizing FM$_p$ layer. After traversing it to become spin-polarized, it  impinges onto the analyzing FM$_a$ layer where it transfers part of its spin angular momentum onto the localized spins via quantum STT~\cite{Zholud2017,Petrovic2021a}. Unlike in conventional  Slonczewski-Berger STT studies~\cite{Ralph2008}---where localized spins within FM$_p$ and FM$_a$ layers are modeled by classical vectors~\cite{Berkov2008}---we retain a fully quantum description based on spin operators. Their ground state expectation values $\langle \hat{\mathbf{S}}_i \rangle(t \le 0)$ are depicted by red arrows and are arranged into a {\em collinear but antiparallel} geometry in which conventional STT is identically zero~\cite{Ralph2008}. This process dynamically generates a mixed entangled  quantum state of  the \mbox{FM$_p \cup$ FM$_a$} subsystem. To account for  its dissipation and phase decoherence, we couple each spin to its own bosonic bath~\cite{Rudner2020,GarciaGaitan2023}, where all such baths are kept in equilibrium at temperature $T_\mathrm{bb}$. Delayed X-ray pulses, with incoming momentum and energy $(\hbar \mathbf{k}_i,\hbar \omega_i)$, are assumed to be shined during and after the current pulse duration. X-rays, with momentum and energy $(\hbar \mathbf{k}_f,\hbar \omega_f)$, scattered off the \mbox{FM$_p \cup$ FM$_a$} subsystem  facilitate a  witnessing scheme of nonequilibrium entanglement we propose based on extraction~\cite{Hales2022} of time-dependent quantum Fisher information from trRIXS response function~\cite{Mitrano2020,Chen2019,Halasz2016}.}
		\label{fig:fig1}
	\end{figure}

	These demonstrations of nonclassical bipartite states generating entanglement between well-separated objects, as well as earlier proposals that exploited radiation pressure effects inside microcavities ~\cite{Ferreira2006,Vitali2007,Clarke2020},  have relied upon the use of photons as {\em mediators} of quantum correlations ~\cite{Marletto2017,Hall2018}. Recently, a greater interest has been placed on trying to reproduce these effects in a solid state setup. In effect, a small $\sim\!\!10$~\cite{Bradeley2020} or moderate $\sim\!\!10^3$~\cite{Kimov2015} number of spins in solids have already been entangled at distances of a few lattice constants, while predictions\,\cite{Venuti_07,Ferreira_08} of long-range entanglement among spin-$1/2$ probes in strongly correlated systems have also been recently realized in experiments with superconducting flux circuits\,\cite{Tennant_22}. Moreover, several recent theoretical works~\cite{Nair2020,Elyasi2020,Yu2020,Ren2022} prescribe a way to entangle much larger spin ensembles ($N \gtrsim 10^{16}$) residing within two {\em distant} spheres carved out of a ferrimagnetic insulator, using the cavity photon modes as entanglement mediators. It is important to note that quantum entanglement of a macroscopic number of degrees of freedom is ubiquitous in the ground or low-lying excited states of strongly electron-correlated materials, such as superconductors~\cite{Vedral2004}, quantum spin liquids~\cite{Grover2013},  antiferromagnets~\cite{Morimae2005,Mathew2020,Scheie2021,Laurell2021,FrancisSong2011}, and Hubbard model materials~\cite{Vafek2017,Hales2022}. However, in practice, it is extremely challenging to isolate different subsystems of such systems and then probe their mutual entanglement.
	
	At first sight, it seems that none of the plethora of {\em nonequilibrium} spin-dependent phenomena, involving itinerant  electrons and localized spins in typical spintronic devices, like spin valves (SVs) and magnetic tunnel junctions (MTJs), would be useful for investigating large-scale entanglement of well-separated quantum units. Anticipating the deception of this expectation, we still recall that SVs and MTJs 
	are composed of two macroscopic FM layers hosting a very large number of localized spins, usually derived from $d$-orbitals  of Fe, Ni, or Co. These FM layers are separated by a few nanometers thick NM spacer (such as Cu) in SVs (as illustrated schematically by our 1D model in Fig.~\ref{fig:fig1}) or by an insulating barrier (such as MgO) in the case of MTJs.  At standard room-temperature, the value of these localized spins within  FMs are typically $S>1$, which falls outside~\cite{Kaxiras2019} of the \textit{``ultra-quantum''} limit, where quantum corrections to the  $S^2(1+1/S)$ eigenvalue of the $\hat{\mathbf{S}}_i^2$ operator are significant~\cite{Parkinson1985}. This fact has been used to intuitively (but not rigorously~\cite{Mondal2021}) justify the modeling of spin dynamics~\cite{Evans2014} and injected electronic currents~\cite{Berkov2008} in the presence of magnetic fields, by means of the Landau-Lifshitz-Gilbert (LLG) equation~\cite{Evans2014,Berkov2008}, which treats localized spins as {\em classical} vectors of fixed length. The extended LLG equation~\cite{Berkov2008}  includes a conventional (Slonczewski-Berger) spin-transfer torque (STT)~\cite{Ralph2008}  term describing  spin angular momentum exchange  at a semiclassical level. Such a term may be phenomenological, as in classical micromagnetics codes~\cite{Berkov2008}, or it can be computed microscopically from some steady-state~\cite{Ellis2017} or time-dependent~\cite{Petrovic2018} {\em single-particle} quantum transport theory. 
	
	Defying conventional wisdom, recent experiments at ultralow temperatures (\mbox{$T \sim 1$ K})~\cite{Zholud2017} have observed current-driven magnetization dynamics in SVs that started from {\em collinear} magnetizations in the two FM layers.\,\,In this situation, the conventional STT is {\em identically zero} and the system's dynamics cannot be understood within the LLG paradigm.~This has motivated the development of a quantum STT theory, where {\em both} the flowing electronic spins and localized spins {\em must} be treated quantum-mechanically~\cite{Petrovic2021a,Mondal2019,Petrovic2021,Mitrofanov2020,Mitrofanov2021}.~Even though a  ferromagnet in equilibrium remains in a separable (unentangled) quantum state  under various externally imposed conditions~\cite{Pratt2004}, a single FM layer can be driven by a spin-polarized current to experience a quantum STT which induces a dynamical build-up of long-range entanglement~\cite{Petrovic2021a,Elben2020}. The simplest signature of such entanglement is a shrinking in the expectation value of the localized spin magnitude, \textit{i.e.},  $|\langle  \hat{\mathbf{S}}_i \rangle (t)|\!<\!S\hbar$ ($i$ being the site of the crystalline lattice). In some circumstances, these expectation values can even be reduced to zero~\cite{Petrovic2021a}, which further explains the failure of the classical LLG equation~\cite{Ralph2008,Berkov2008} to describe the STT-driven magnetization dynamics on these systems~\cite{Mondal2021,Wieser2015}. The quantum nature of the problem then calls for a full-fledged many-body approach that captures the intrinsic quantum nature of localized spins and thus goes beyond the common paradigm of classical magnetization dynamics.

	Here, we exploit the quantum STT mechanism as a means to entangle two {\em distant} FM layers of a nonequilibrium SV device that is achievable by modern techniques of nanofabrication (depicted in Fig.~\ref{fig:fig1}). The  quantum STT is driven by a  {\em spin-unpolarized} electronic current pulse that is injected into the NM and plays the role of an entanglement mediator.~As the pulse travels through the device, it first becomes (partially) spin-polarized by interacting with the polarizing FM layer (FM$_p$) and then, after traversing the spacer, it eventually exchanges spin angular momentum via quantum STT with the analyzing FM layer (FM$_a$).~As required by general theorems~\cite{Marletto2017,Hall2018}, the mediating pulse must be intrinsically quantum mechanical, which assumes a sufficiently low working temperature~\cite{Zholud2017} and requires that decoherence mechanisms affecting the charge-spin dynamics are suppressed during its traveling time across the entire device.~Under these conditions, we predict that both FM layers will become mutually entangled over time, to a degree that can be easily controlled by parameters such as the magnitude and duration of injected current pulse.~Furthermore, as the device operates, the FM$_p$$\cup$FM$_a$ subsystem also becomes entangled with the mediating pulse, placing the former into a {\em mixed and entangled} bipartite quantum state. 
	
	The paper is organized as follows. Section~\ref{sec:mln} overviews the   measures of mixed-state entanglement employed in this study,  while Sec.~\ref{sec:models} introduces useful concepts and notation, including the second-quantized Hamiltonian as a microscopic model of our SV device and the many-body algorithms employed in this work. The results for single-electron current pulse as mediator of entanglement, including effects due to thermal fluctuations and coupling to bosonic baths, are reported in Secs.~\ref{sec:single} and ~\ref{sec:tomography}; while zero-temperature results for a many-electron pulse as mediator of entanglement are analyzed in Sec.~\ref{sec:multi}. In Sec.~\ref{sec:witnessing} we propose  an experimental scheme for witnessing macroscopic and nonequilibrium entanglement of two FM layers based on application~\cite{Hales2022} of the state-of-the-art time-resolved resonant inelastic X-ray scattering (trRIXS) technique~\cite{Mitrano2020,Chen2019,Halasz2016}. We conclude in Sec.~\ref{sec:conclusions}. 
	
	\section{Mixed entangled states and measures of their entanglement}\label{sec:mln}
	
	The quantum state of localized spins of the FM$_p$$\cup$FM$_a$ subsystem will become a mixed entangled one in the course of time evolution (Figs.~\ref{fig:fig3}-\ref{fig:fig3size}). That is, it will be described by a reduced density matrix which is {\em not}  expressible as a convex combination of direct product states 
	\begin{equation}\label{eq:mixedentangled}
		\hat{\rho}_{\mathrm{FM}_p \cup \mathrm{FM}_a} \neq \sum_i p_i \hat{\rho}_{\mathrm{FM}_p}^i \otimes \hat{\rho}_{\mathrm{FM}_a}^i,
	\end{equation}
	acting in the bipartite Hilbert space $\mathcal{H}_{\mathrm{FM}_p} \otimes \mathcal{H}_{\mathrm{FM}_a}$. Here $\hat{\rho}_{\mathrm{FM}_p \cup \mathrm{FM}_a} =  \mathrm{Tr}_e \hat{\rho}(t)$, where $\hat{\rho}(t)$ is density matrix of the total system  FM$_p$$\cup$FM$_a$$\cup$electrons and the partial trace is performed over the degrees of freedom of the itinerant electrons. Moreover, $\hat{\rho}_{\mathrm{FM}_p}$ and  $\hat{\rho}_{\mathrm{FM}_a}$ are analogously defined density matrices of the individual layers FM$_p$ and FM$_a$. An equality in Eq.~\eqref{eq:mixedentangled} would signify  separable (unentangled) mixed quantum state~\cite{Peres1996,Wu2020,Elben2020a,Murciano2022,Zhou2020,Sang2021}. Although enormous progress has been made in the last two decades in understanding entanglement of pure quantum many-body states~\cite{Chiara2018,Laflorencie2016,Beneti2019}, much less is understood regarding the nature of quantum correlations in interacting  quantum many-body systems in mixed states. Thus, how to detect and quantify mixed-state entanglement in quantum devices containing many interacting particles is a topic of great and emerging interest~\cite{Wu2020,Elben2020a,Zhou2020}. The mixed states arise due to thermal fluctuations~\cite{Wu2020}, due to decoherence by an external environment or because they describe a subsystem of interest within a much larger and globally entangled system described by a pure state.
	
	To describe the entanglement of a quantum\,many-body mixed\,state $\hat{\rho}_{\mathrm{FM}_p \cup \mathrm{FM}_a}$ of a FM$_a$$\cup$FM$_p$ subsystem, we calculate the {\em mutual logarithmic negativity} (MLN) between the $\mathrm{FM}_p$ and $\mathrm{FM}_a$ layers defined by~\cite{Sang2021}
	\begin{eqnarray}\label{eq:negativity}
		E_{N}({\mathrm{FM}_p|\mathrm{FM}_a}) & \equiv & E_{N}(\hat{\rho}_{\mathrm{FM}_p \cup \mathrm{FM}_a}) = \ln || \hat{\rho}^{T_{\mathrm{FM}_p}}_{\mathrm{FM}_p \cup \mathrm{FM}_a} ||_1   \nonumber \\
		&=& \ln || \hat{\rho}^{T_{\mathrm{FM}_a}}_{\mathrm{FM}_p \cup \mathrm{FM}_a} ||_1 = \ln \sum_n |\lambda_n|,
	\end{eqnarray}
	where  $||\hat{A}||_1 = \mathrm{Tr}|\hat{A}|=\mathrm{Tr} \sqrt{\hat{A}^\dagger \hat{A}}$ is the trace norm of operator $\hat{A}$; and $\lambda_n$ are the eigenvalues of $\hat{\rho}^{T_{\mathrm{FM}_p}}_{\mathrm{FM}_p \cup \mathrm{FM}_a}$ or $\hat{\rho}^{T_{\mathrm{FM}_a}}_{\mathrm{FM}_p \cup \mathrm{FM}_a}$. 
	The matrix elements of the partial transpose with respect to, e.g., FM$_p$ are given by~\cite{Beneti2019}
	\begin{equation}\label{eq:partialtranspose}
		\big(\hat{\rho}^{T_{\mathrm{FM}_p}}_{\mathrm{FM}_p \cup \mathrm{FM}_a} \big)_{i\alpha;j\beta} = \big(\hat{\rho}_{\mathrm{FM}_p \cup \mathrm{FM}_a} \big)_{j\alpha;i\beta},
	\end{equation}
	using matrix elements of $\hat{\rho}_{\mathrm{FM}_p \cup \mathrm{FM}_a}$
	\begin{equation}\label{eq:matrixelements}
		\big(\hat{\rho}_{\mathrm{FM}_p \cup \mathrm{FM}_a} \big)_{i\alpha; j\beta} = {}_{\mathrm{FM}_p} \langle i| \, {}_{\mathrm{FM}_a}\langle \alpha|\hat{\rho}_{\mathrm{FM}_p \cup \mathrm{FM}_a}| j \rangle_{\mathrm{FM}_p} |\beta \rangle_{\mathrm{FM}_a}.
	\end{equation}
	Although the MLN can be zero for an entangled mixed state, a nonzero MLN necessarily implies existence of entanglement between the two parts. The MLN also offers a useful probe for distinguishing bipartite and multipartite quantum correlations in a pure state of the total system---for tripartite pure state $|\Psi \rangle_{\mathrm{FM}_p \cup \mathrm{FM}_a \cup e}$ of  a system composed of all localized spins and injected electrons, as denoted by FM$_p$$\cup$FM$_a$$\cup$electrons, MLN of $\hat{\rho}_{\mathrm{FM}_p \cup \mathrm{FM}_a}$ in Eq.~\eqref{eq:negativity} detects genuine quantum correlations between FM$_a$ and FM$_p$. In the SV device depicted in Fig.~\ref{fig:fig1}, they build-up dynamically [Figs.~\ref{fig:fig3}(d) and Fig.~\ref{fig:fig3size}] for $t > 0$ when the device is out of equilibrium, while not being present prior ($t \le 0$) to the injection of current pulse  when the SV remains in equilibrium. 
	
	In addition, we also use the mutual information (MI) [Fig.~\ref{fig:fig3}(c)] between $\mathrm{FM}_p$ and $\mathrm{FM}_a$ layers
	\begin{equation}\label{eq:mutual}
		\mathcal{I}(\mathrm{FM}_p|\mathrm{FM}_a) = \mathcal{S}_{\mathrm{FM}_p} + \mathcal{S}_{\mathrm{FM}_a} - \mathcal{S}_{\mathrm{FM}_p \cup \mathrm{FM}_a},
	\end{equation}
	obtained from the standard~\cite{Chiara2018,Laflorencie2016} von Neumann entanglement entropy [Figs.~\ref{fig:fig3}(a) and ~\ref{fig:fig3}(b)]
	\begin{equation}\label{eq:entropy}
		\mathcal{S}_\mathrm{sub}(t) = -\mathrm{Tr}\, \left[ \hat{\rho}_\mathrm{sub}(t) \ln \hat{\rho}_\mathrm{sub}(t) \right],
	\end{equation}
	where $\hat{\rho}_\mathrm{sub}(t)$ is the time-dependent nonequilibrium reduced density matrix  of a chosen subsystem (sub), such as sub=FM$_p$$\cup$FM$_a$,  sub=FM$_p$ and sub=FM$_a$. The MI~\cite{Sang2021,Chiara2018,Laflorencie2016} is sensitive to both quantum and classical correlations that arise from tripartite correlations between FM$_p$, FM$_a$ and all injected electrons, unlike the MLN which detects only genuine quantum correlations~\cite{Beneti2019,Groisamn2005}.
	
	For a single injected electron ($N_e=1$), the reduced density matrices  $\hat{\rho}_\mathrm{sub}(t)$ are obtained by directly solving the von Neumann or the Lindblad equation for the density matrix of the total system FM$_p$$\cup$FM$_a$$\cup$electrons and then partially tracing over electronic quantum states. For a multi-electron ($N_e>1$) current pulse, we use the adaptive time-dependent density matrix renormalization (tDMRG) algorithm~\cite{White2004,Schmitteckert2004,Daley2004,Feiguin2011,Paeckel2019}, tailored for quantum STT~\cite{Petrovic2021a,Petrovic2021}. In the latter case we calculate  the von Neumann entropy of the density matrix of {\em half} of the lattice, $\mathcal{S}_\mathrm{half}(t)$, a measure often employed in studies of entanglement of quantum many-body systems~\cite{Petrovic2021a,Bardarson2012}.

	\vspace{-0.4cm}	
	\section{Models and methods}\label{sec:models}
	
	The dynamics of the 1D SV device in Fig.~\ref{fig:fig1} is described by the quantum many-body Hamiltonian
	\begin{eqnarray}\label{eq:hamiltonian}
		\nonumber \hat{H} &=& -\gamma\sum_{\langle ij \rangle}\hat{c}^\dagger_{i\sigma}\hat{c}_{j\sigma} -J_\mathrm{sd} \sum_i  \hat{\mathbf{s}}_i \cdot \hat{\bf S}_i \nonumber \\
		\mbox{}&&-\sum_{\langle ij \rangle}\left[ J\left(\hat{S}_i^x\cdot\hat{S}_j^x+ \hat{S}_i^y\cdot\hat{S}_j^y\right) + J_z\hat{S}_i^z\cdot\hat{S}_j^z\right],
	\end{eqnarray}
	where $i$ and $j$ label positions in the chain and $\langle ij \rangle$ signifies that couplings exists only between the  nearest-neighbor (NN) sites.~As customary, $\hat{c}^\dagger_{i\sigma}$ ($\hat{c}_{i\sigma}$) is a fermionic creation (annihilation) operator for an electron with spin $\sigma=\uparrow,\downarrow$ placed on site $i$ of an \textit{s}-orbital tight-binding chain of $L_x=201$ sites with NN hopping between them of strength \mbox{$\gamma=1$ eV}. This sets an energy scale in the problem, as well as a unit of velocity for the propagating electrons. Also in Eq.~\eqref{eq:hamiltonian}, the second term describes an on-site $sd$ exchange interaction of strength \mbox{$J_\mathrm{sd}=0.5$ eV}~\cite{Cooper1967} between the electronic spin and a localized Heisenberg spin. The last term is an NN Heisenberg XXZ ferromagnetic Hamiltonian that describes localized spins within $\textrm{FM}_p$ and $\textrm{FM}_a$ layers with the $z$-axis as the easy axis. 
	
	To describe the spin of the propagating electron, we define the operator 
	\begin{equation}\label{eq:spinpersite}
		\hat{s}^{\alpha}_{i}=\!\!\!\!\sum_{\sigma=\{\uparrow,\downarrow\}}\!\!\! \hat{c}^\dagger_{i\sigma} \hat{\sigma}^\alpha_{\sigma\sigma'} \hat{c}_{i\sigma'},
	\end{equation}
	as the $\alpha$-component of the electronic spin density in site $i$, where $\hat{\sigma}^\alpha$ is a Pauli matrix with $\alpha \in \{ x,y,z \}$. In this way, the sum of $\hat{s}_{i}^{\alpha}$ over $i$ yields the total electronic spin operator
	\begin{equation}\label{eq:totalspin}
		\hat{s}_{e}^{\alpha} =  \sum_i \hat{s}^{\alpha}_{i}.
	\end{equation}
	By contrast, each localized spin-$\frac{1}{2}$ has a two-dimensional Hilbert space, $\mathcal{H}_n$ ($n=1,...,N$), so that in Eq.~\eqref{eq:hamiltonian} 
	the localized spin at site $i$ is described by an operator \mbox{$\hat{S}_i^\alpha = \underbrace{\hat{I} \otimes \hat{I} \cdots \hat{I}}_\text{$i-1$} \otimes \hat{\sigma}^\alpha \underbrace{\otimes \hat{I} \otimes \cdots \otimes \hat{I}}_\text{$N-i$ times}$}, where $\hat{I}$ is the unit operator (or unit $2 \times 2$ matrix in representation). The localized spin interacts with its NNs within the same FM layer 
	by an XXZ Heisenberg term of strength \mbox{$J = 0.1$ eV} and \mbox{$J_z = 1.005J$}. The total number of localized spins is $N$, half of which  belong to FM$_p$ layer and the other half to FM$_a$ layer. The two layers are separated by $N_\mathrm{NM}$ sites with no localized spins so as to simulate the NM layer depicted in Fig.~\ref{fig:fig1}.

	Thus, our model describes the FM layers as quantum Heisenberg chains which, in place of being insulating, are effectively metallic as there are electrons hopping between their sites.~From the viewpoint of strongly correlated electron physics, each FM layer can be interpreted as a Kondo-Heisenberg model~\cite{Tsvelik2017} sandwiched by fermionic leads (i.e., the chain sites hosting no localized spins), with a ferromagnetic ($J,J_z>0$) exchange interaction between localized spins as well as with the spin of the flowing electrons ($J_\mathrm{sd}>0$).
	
	We will consider in the remainder of this paper that the chain of $L_x$ sites is traversed by either (\textit{i}) a single (spin-unpolarized) electron pulse, or (\textit{ii}) a pulse composed  of equal number of spin-up and spin-down electrons. For a single injected electron ($N_e=1$), the system has support in the Hilbert space  \mbox{$\mathcal{H} = \mathcal{H}^\text{orb}_e \otimes  \mathcal{H}^\text{spin}_e \otimes \mathcal{H}_1 \otimes \ldots \otimes \mathcal{H}_N$}, where $\mathcal{H}^\text{orb}_e$ ($\mathcal{H}^\text{spin}_e$) is the space of orbital (spin) states of the electron. In the multi-electron simulations, even number $N_e>1$  of electrons are injected in the chain, so that  the system  Hamiltonian now acts in the space $\mathcal{H} = \mathcal{F}_e \otimes \mathcal{H}_1\otimes\cdots\otimes\mathcal{H}_N$ where $\mathcal{F}_e$ is the Fock space for many spin-$\frac{1}{2}$ itinerant electrons. 
	
	\vspace{-0.4cm}
	\subsection{Unitary time-evolution for an injected single-electron pulse (via the von Neumann equation)}\label{sec:vonneumann}
	
	We start by considering that a single spin-unpolarized and right-propagating electron pulse is injected into the fermionic chain, from the left~\cite{Mondal2019,Mitrofanov2020,Mitrofanov2021,Kim2007,Dogan2009}.~Such situation can be experimentally realized~\cite{Dubois2013} by applying a Lorentzian voltage pulse that excites a \textit{soliton-like} quasiparticle---the so-called leviton~\cite{Keeling2006}---of elementary charge (\mbox{$\,Q\!=\!\!\int \! I(t) dt\!=\! e\,$}), out of the Fermi sea. Therefore, the initial quantum state of the complete system, at zero temperature, is described by the density matrix 
	\begin{eqnarray}\label{eq:initialdm}
		\hat{\rho}(t=0) & \equiv &  |\Phi \rangle\! \langle \Phi| \otimes \frac{1}{2} \big(|\!\uparrow\rangle\! \langle \uparrow\!| + |\!\downarrow \rangle \!\langle \downarrow\!| \big) \nonumber \\
		&& \otimes | \mathrm{FM}_p \rangle \!\langle \mathrm{FM}_p| \otimes |\mathrm{FM}_a \rangle\! \langle \mathrm{FM}_a|,
	\end{eqnarray}
	where the two leading terms refer to the quantum state of propagating electron, and the last are oppositely magnetized states of both FM layers, i.e., $\!| \mathrm{FM}_p \rangle \!=\! |\!\! \uparrow \uparrow \uparrow \uparrow \rangle \!\in \!\mathcal{H}_1 \otimes \ldots \otimes \mathcal{H}_{N/2}$; and $| \mathrm{FM}_a \rangle = |\!\! \downarrow \downarrow \downarrow \downarrow \rangle \in \mathcal{H}_{N/2+1} \otimes \ldots \otimes \mathcal{H}_N$. Note that the state in Eq.~\eqref{eq:initialdm} is an example of a multi-partite but fully separable (unentangled) mixed state~\cite{Wu2020,Elben2020a} which is almost pure, except for the second factor that expresses the zero spin-polarization of the injected electron pulse. The corresponding orbital state is pure and given by the Gaussian wavefunction 
	\begin{equation}\label{eq:wp}
		\langle x|\Phi \rangle= C\exp{\left(ik_x x-\delta_{k_x}^2(x-x_0)^2/4\right)}, 
	\end{equation}
	with $C$ as the normalization constant, and $k_x=\pi/2$ and $\delta_{k_x}=0.2$ selected to reduce dispersive spreading~\cite{Goussev2018} of such quantum wavepacket. The propagation of the wavepacket through SV in Fig.~\ref{fig:fig1} is animated in the movie provided as the Supplemental Material (SM)~\cite{sm}.   
	
	Since entanglement in solids inevitably decays as temperature increases~\cite{Brukner2006,Mathew2020,Scheie2021,Laurell2021}, we also consider the time evolution of the system at finite temperature by switching the initial quantum state from Eq.~\eqref{eq:initialdm} to
	\begin{eqnarray}\label{eq:initialdmfinitet}
		\hat{\rho}(t\!=\!0) & \!\equiv\! &  |\Phi \rangle\! \langle \Phi| \!\otimes\! \frac{1}{2} \big(|\!\uparrow\rangle\! \langle \uparrow\!|\! +\! |\!\downarrow \rangle\! \langle \downarrow\!| \big) \!\otimes\! \rho^\mathrm{eq}_{\mathrm{FM}_p \cup \mathrm{FM}_a},
	\end{eqnarray}
	where $\rho^\mathrm{eq}_{\mathrm{FM}_p \cup \mathrm{FM}_a}\!\!\! =\!\! \mathcal{Z}^{\scriptscriptstyle{-1}}\exp \big( {-\beta H_{\mathrm{FM}_p \cup \mathrm{FM}_a}} \big)$ is the equilibrium density matrix of all localized spins in the canonical ensemble at temperature $T\!\!=\!\!1/\beta k_{\text{B}}$,  and \mbox{$\mathcal{Z}=\mathrm{Tr}\, \exp\big(-\beta H_{\mathrm{FM}_p \cup \mathrm{FM}_a} \big)$} is the corresponding partition function. 
	
	In both cases, the unitary time-evolution of $\hat{\rho}(t)$ is computed by solving the von Neumann equation
	\begin{equation}\label{eq:vnm}
		i\hbar \frac{\partial \hat{\rho}}{\partial t} = \left[\hat{H}, \hat{\rho} \right].
	\end{equation} 
	where two algorithms are used. To evolve the zero-temperature state, we employ a  spectral method based upon an efficient and stable Chebyshev polynomial expansion of the time-evolution operator~\cite{TalEzer84,Pires2020}. This technique, coupled with a robust parallelization scheme~\cite{Joao2020}, allow us to simulate large tight-binding lattices with FM layers that host up to 10  localized spins each (labeled as $10+10$ case in Fig.~\ref{fig:fig6}). For the finite temperature time-evolution of $4+4$ localized spins [Fig.~\ref{fig:fig3}(d)], we used standard 4$^\mathrm{th}$-order Runge-Kutta method converged with a time step of \mbox{$\delta t = 0.1$ fs}.
	
	\vspace{-0.5cm}
	
	\subsection{Nonunitary time-evolution for an injected single-electron pulse plus decoherence (via the Lindblad equation)}\label{sec:lindblad} 
	
	In a realistic SV device,  the degrees of freedom in question (electron pulse + FM layers) form an open quantum system subject to {\em dissipation and phase decoherence}.~To investigate the impact of such processes, we consider nonunitary time-evolution due to localized spins being coupled to a generic heat bath of bosons in thermal equilibrium at temperature $T$. We assume that there are no significant correlations between  the SV system and the bath at $t=0$ and that the time scale of evolution of the SV is much greater than the bath's correlation time~\cite{deVega2017}. Such a Markovian time-evolution, wherein memory effects are suppressed, greatly simplifies matters and will allow us to assess how the entanglement generation is affected by coupling to a dissipative  environment.

	For this purpose, the Hamiltonian  [Eq.~\eqref{eq:hamiltonian}] is supplemented with two additional terms 
	\begin{equation}\label{eq:hplusbath}
		\hat{H}_{\text{tot}}= \hat{H}+\hat{H}_{\text{bath}} + \hat{V},
	\end{equation}
	Here $\hat{H}_{\text{bath}}$ is the Hamiltonian of the bosonic bath, modeled as a set of harmonic oscillators
	\begin{equation}\label{eq:bath}
		\hat{H}_{\text{bath}}= \sum_{ik} w_{ik} \hat{a}_{ik}^\dagger \hat{a}_{ik},
	\end{equation}
	with an operator $\hat{a}_{ik}$ ($\hat{a}_{ik}^\dagger$)  annihilating (creating) a boson in mode $k$. This model interacts with spin operator at site $i$~\cite{Breuer2007} via 
	\begin{equation}\label{eq:coupling}
		\hat{V} = \sum_k g_k \sum_i  \hat{\mathbf{S}}_{i}  (\hat{a}_{ik} + \hat{a}_{ik}^\dagger),
	\end{equation}
	where $g_k=0.12$ are the coupling constants. We assume that each spin interacts with the bosonic bath independently of other spins~\cite{Rudner2020,GarciaGaitan2023}. Furthermore, if we assume small $g_k$, the quantum master equation (QME) of the Lindblad type~\cite{Lindblad1976,Manzano2020} can be derived by tracing out the bosonic bath and by expanding the resulting equation to second order. 
	
	The study of quantum spins interacting with bosonic bath has a long history, dating back to spin-boson model~\cite{Leggett1987} and its generalizations~\cite{Anders2022}, as archetypal systems for exploring  dissipative quantum mechanics.  Furthermore, for problems involving magnetic materials hosting many spins~\cite{GarciaGaitan2023,Norambuena2020}, or for multiple qubits~\cite{Zou2022}, one needs to go beyond one quantum spin $S=1/2$ of the spin-boson model. Such extensions have been pursued in recent derivations of QMEs for two~\cite{Zou2022} or more~\cite{Norambuena2020} quantum spins interacting both with their NNs and with external bosonic bath. However,  traditional approaches for the derivation of the Lindblad QME---such as using Born, Markov and secular approximations~\cite{Manzano2020,Schaller2014}---work well for two spins~\cite{Zou2022} but fail already for four spins~\cite{GarciaGaitan2023} because of (nearly) degenerate eigenlevels. To avoid this problem, we follow the procedure of Ref.~\cite{Rudner2020} for deriving an universal Lindblad QME which evades  difficulties of the secular approximation~\cite{Schaller2016}. This evades limitations in recently derived~\cite{Norambuena2020} Lindblad QMEs for many localized quantums spins due to unwarranted assumptions made on the relative size of energy splittings compared to the bath fluctuations. The universal Lindblad QME~\cite{Rudner2020} considers a single Lindblad operator $\hat{L}_i$ for each spin, so that $N$ such operators are needed to obtain
	\begin{equation}
		\label{eq:ULE}
		d{\hat{\rho}}/dt = -i [\hat{H},\hat{\rho}] + \sum_i^N \hat{L}_i \hat{\rho} \hat{L}^\dagger_i-\frac{1}{2}\{\hat{L}_i^\dagger \hat{L}_i, \hat{\rho} \},
	\end{equation}
	where we also ignore typically negligible Lamb-shift corrections~\cite{Schaller2014} to the Hamiltonian. The Lindblad QME is time-local due to the assumption that bath-induced changes to the system dynamics are slow relative to the typical correlation time of the bath. Following Ref.~\cite{GarciaGaitan2023}, we compute $\hat{L}_i$ operators as a power series
	\begin{equation}
		\label{eq:expand}
		\hat{L}_i = \sum_n c_n (\text{ad}_{\hat{H}})^n [\hat{\mathbf{S}}_i], \: \: \: \: c_n=\frac{(-i)^n}{n!}\int_{-\infty}^\infty dt g(t) t^n,
	\end{equation}
	where the sum is truncated to $n=$1--8 terms. Here $\text{ad}_{\hat{H}} [X] = [\hat{H},X]$ and the jump correlator function is defined via the Fourier transform of the spectral function of the bath, \mbox{$J(\omega)=2\pi \sum \delta(\omega-\omega_k)$}, as
	\begin{equation}
		\label{eq:jumpcorr}
		g(t)=\frac{1}{\sqrt{2\pi}}\int_{-\infty}^{\infty} d\omega\sqrt{J(\omega)} e^{-i \omega t}.
	\end{equation}
	For numerical calculations, we considered an Ohmic spectral function with a rigid ultraviolet cutoff
	\begin{equation} \label{eq:Spectral}
		J(\omega) = \Gamma \omega / \omega_m \Theta(\omega_m-\omega) n_{\mathrm{BE}}(\omega),
	\end{equation}
	where $\Gamma$ is the reorganization energy representing the magnitude of fluctuations and dissipation; $\omega_m$ characterizes how quickly the bath relaxes towards equilibrium; $n_\mathrm{BE}(\omega)$ the Bose-Einstein distribution; and $\Theta$ is the Heaviside step function.  The Lindblad QME [Eq.~\eqref{eq:ULE}] is only valid for a weak spin-bath coupling, as it assumes a second order truncation in $g_k$. 
	
	We do not directly couple the itinerant electrons to a dissipative environment, as it is common in calculations of conventional STT~\cite{Ralph2008} where current of injected electrons is always treated by some version of fully phase-coherent quantum transport formalism~\cite{Ralph2008,Ellis2017,Wang2008,Dolui2020}. The reason is that STT is an interfacial phenomenon, where spin angular momentum is absorbed within few tens of lattice spacings away from the NM/FM$_a$ interface~\cite{Wang2008}, so that such length scale is  shorter than dephasing length for charge and spin degrees of freedom of itinerant electrons (even at room temperature). We solve Eqs.~\eqref{eq:ULE} and ~\eqref{eq:expand} for the system of a single coherent electron $+$ localized spins coupled to bosonic baths, as illustrated in Fig.~\ref{fig:fig1}, using quantum trajectories algorithm~\cite{Daley2014} where we average over 200  trajectories. 
	
	\vspace{-0.4cm}
	\subsection{Unitary time-evolution for an injected multi-electron pulse (via tDMRG)}\label{sec:tdmrg}
	
	Unlike the single-electron pulse cases, the time-evolution of the system upon the injection of a multi-electron pulse is handled using a tDMRG algorithm~\cite{White2004,Schmitteckert2004,Daley2004,Feiguin2011,Paeckel2019,Petrovic2021a,Petrovic2021}. This requires the following extra terms to be included into the Hamiltonian of Eq.~\eqref{eq:hamiltonian}   
	\vspace{-0.3cm}
	\begin{align}\label{eq:hamiltonianconfine}
		\hat{H}_\mathrm{V,\mathbf{B}} =&  -V(t\! \le\! 0)\! \sum_{i=1}^{N_\mathrm{conf}}\!\!
		\left(
		\hat{c}^\dagger_{i\uparrow}\hat{c}_{i\uparrow}
		+
		\hat{c}^\dagger_{i\downarrow}\hat{c}_{i\downarrow}
		\right) \\
		&\qquad+g \mu_B B(t \le 0)\left(\sum_{i \in \mathrm{FM}_a} 
		\!\!\! \hat{S}_{i}^z
		\!-\!\!\!\!\sum_{i \in \mathrm{FM}_p} 
		\!\!\!\hat{S}_{i}^z\right)\nonumber,
	\end{align}
	where $g$ is the electron  gyromagnetic ratio and $\mu_B$ is the Bohr magneton. The first term on the right-hand side (RHS) is a scalar potential term ($V\!=\!2 \gamma$) that acts only for times $t\! \le\! 0$, in order to confine all electrons within $N_\mathrm{conf}\!=\!10$ sites at the left edge of Fig.~\ref{fig:fig1}.~Because the potential acts equally on both spin sectors,  confined electrons remain spin-unpolarized, i.e., with their total spin $\langle \hat{\mathbf{s}}_e \rangle\!(t \le 0) =\!0$.~For \mbox{$t\! >\! 0$}, the confinement is switched off and the electrons get injected into FM$_p$/NM/FM$_a$ region of Fig.~\ref{fig:fig1}, spreading from left to right. Likewise, the last terms on the RHS of Eq.~\eqref{eq:hamiltonianconfine} introduce an external magnetic field $B(t\! \le\! 0)\!=\!10\gamma/g \mu_B$ at negative times, which polarizes the localized spins of the FM$_p$ (FM$_a$) layer along the $+z$-axis ($-z$-axis), guaranteeing that $\langle \hat{S}_i^z \rangle/ S\hbar\! =\! +1$ ($\langle \hat{S}_i^z \rangle/S\hbar\! =\! -1$) at $t\!=\!0$ as illustrated in Fig.~\ref{fig:fig1}. In this way, the ground state of Eq.~\eqref{eq:hamiltonianconfine} properly initializes the whole system as a quantum many-body state $|\Psi(t\!=\!0) \rangle_{\mathrm{FM}_p \cup \mathrm{FM}_a \cup e}$ at the beginning of the time evolution.  

	\begin{figure}[!t]
		\includegraphics[width=\linewidth]{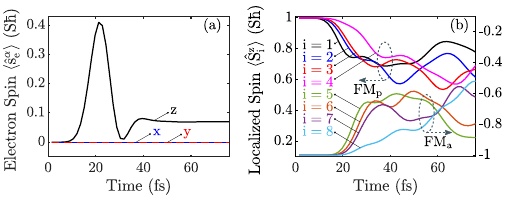}
		\vspace{-0.6cm}
		\caption{Time evolution of the expectation values of (a) flowing electronic spin and (b) localized spins after injection of a  single-electron ($N_e=1$) current pulse that is spin-unpolarized at  $t=0$. The  initial quantum state of the whole system, with antiparallel magnetizations of the FM$_p$ and FM$_a$ layers of the SV in Fig.~\ref{fig:fig1}, is  described by the density matrix in Eq.~\eqref{eq:initialdm} at $T=0$ K. In panel (b), localized spins $i=1$--$4$ belong to the FM$_p$ layer, and $i=5$--$8$ belong to the FM$_a$ layer.}
		\vspace{-0.55cm}
		\label{fig:fig2}
	\end{figure}

	The adaptive tDMRG algorithm~\cite{White2004,Schmitteckert2004,Daley2004,Feiguin2011,Paeckel2019} evolves the nonequilibrium state of the total system
	\begin{equation}\label{eq:evolutionoperator}
		|\Psi(t+\delta t) \rangle_{\mathrm{FM}_p \cup \mathrm{FM}_a \cup e} = e^{-i\hat{H} \delta t/\hbar} |\Psi(t)\rangle_{\mathrm{FM}_p \cup \mathrm{FM}_a \cup e},
	\end{equation} 
	using the time step \mbox{$\delta t =0.1 \hbar/\gamma$}. We start the propagation with $m = 100$ states and limit the truncation error to $10^{-7}$, while the maximal number of states allowed during the evolution is set to $m_\mathrm{max} = 400$.  Since the simulated chains are finite (unlike standard single-particle quantum transport calculations with semi-infinite leads~\cite{Petrovic2018}), the SV system can be evolved only for a limited amount of time (Fig.~\ref{fig:fig4}) before electrons are backscattered by the right boundary, breaking  the left$\rightarrow$right current flow. Nevertheless, the quantum dynamics of the injected electrons is effectively equivalent to the dynamics in an infinite~\cite{Petrovic2018,Eckel2010} open quantum system before the boundary reflection takes place.

	\section{Results and Discussion}\label{sec:results}
	
	\subsection{Single-electron pulse injection}\label{sec:single}

	\begin{figure}[!t]
		\includegraphics[width=8.5cm]{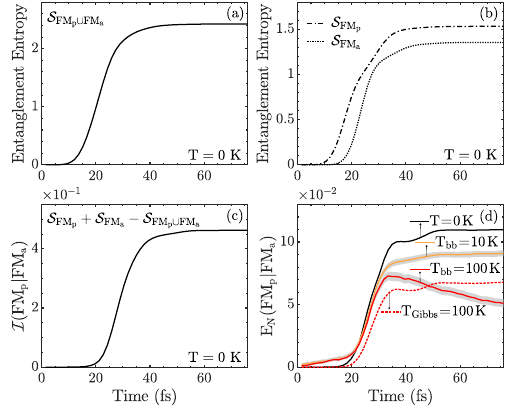}
		\vspace{-0.2cm}
		\caption{Time evolution of the von Neumann entropy [Eq.~\eqref{eq:entropy}] of localized spins at $T=0$ K within (a) FM$_p$$\cup$FM$_a$; and (b) FM$_p$ (dash-dotted line) or FM$_a$ (dotted line) subsystems of SV in Fig.~\ref{fig:fig1}. From the entropies in (a) and (b), we construct in panel (c) the MI between FM$_p$ and FM$_a$ layers. Panel (d) plots the corresponding MLN [Eq.~\eqref{eq:negativity}] between the FM$_p$ and FM$_a$ layers at \mbox{$T=0$ K} (solid black line). For comparison, panel (d) also plots the MLN at temperature \mbox{$T_\mathrm{Gibbs}=100$ K} (dotted red line) using initial condition in Eq.~\eqref{eq:initialdmfinitet} where localized spins are described by the Gibbs canonical ensemble density matrix at $t=0$; as well as in the presence of bosonic baths~\cite{Rudner2020,GarciaGaitan2023} interacting [Fig.~\ref{fig:fig1}] with localized spins (solid orange and red lines for \mbox{$T_\mathrm{bb}=10$ K} and \mbox{$T_\mathrm{bb}=100$ K}, respectively) and evolved via Eqs.~\eqref{eq:ULE} and ~\eqref{eq:expand} while using initial condition in Eq.~\eqref{eq:initialdm}. The gray strip around the solid orange and red lines quantifies statistical errors in quantum trajectories algorithm~\cite{Daley2014} (where averaging over 200 trajectories is performed) for solving the Lindblad QME [Sec.~\ref{sec:lindblad}].}
		\label{fig:fig3}
		\vspace{-0.5cm}
	\end{figure}

	We first present the results obtained from evolving the SV device upon the injection of a single-electron pulse. Figure~\ref{fig:fig2} shows the expectation value of the electronic spin, as well as of all $4+4$ localized spins, over time. The results show that the electronic pulse starts from a spin-unpolarized state that develops a finite $z$-axis polarization, $\langle \hat{s}_e^z\rangle$, as it passes through the FM$_p$ layer. This quantum average is obtained as 
	\begin{equation}\label{eq:rhoe}
		\langle \hat{\mathbf{s}}_e \rangle(t) = \mathrm{Tr}\, \big[\hat{\rho}^\mathrm{spin}_e(t) \hat{\bm \sigma} \big],
	\end{equation}
	where $\hat{\rho}_e^\mathrm{spin} = \mathrm{Tr}_\mathrm{other} \hat{\rho}(t)$ with the partial trace performed over the states in $\mathcal{H}^\text{orb}_e  \otimes \mathcal{H}_1 \otimes \ldots \otimes \mathcal{H}_N$. As the electron pulse becomes polarized, the dynamics of the localized spins within the FM$_p$ layer is also initiated [top bundle of curves in Fig.~\ref{fig:fig2}(b)] at \mbox{$t \simeq 10$ fs}.  The partially polarized (by the FM$_p$ layer) spin of the injected electron starts to decay  at around \mbox{$t \simeq 20$ fs} in Fig.~\ref{fig:fig2}(a), which is exactly the time when the localized spins  within the FM$_a$ layer [bottom bundle of curves in Fig.~\ref{fig:fig2}(b)] start to evolve due to  transfer~\cite{Petrovic2021a} of spin angular momentum. The signature of such quantum-STT-driven localized spin dynamics within the FM$_a$ layer is the evolution of only the $\langle \hat{S}_i^z \rangle(t)$ component, without any rotation away from the $z$-axis because $\langle \hat{S}_i^x \rangle(t) = \langle \hat{S}_i^y \rangle(t) \equiv 0$. This is a {\em genuinely  nonclassical} magnetization dynamics that is impossible within the framework of the LLG equation where STT-driven $\langle \hat{\mathbf{S}_i} \rangle(t)$ can only rotate while 
	keeping its length constant~\cite{Ralph2008,Berkov2008}. 

	\begin{figure}[!t]
		\includegraphics[width=\linewidth]{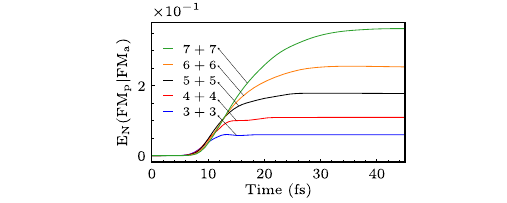}
		\caption{Time evolution of the MLN between the FM$_p$ and FM$_a$ layers  at \mbox{$T=0$ K} for the same single-electron ($N_e=1$) spin-unpolarized current pulse injection used in Figs.~\ref{fig:fig2} and ~\ref{fig:fig3}, but with increasing number of localized spins 
			$N/2 + N/2$ within FM$_p$ and FM$_p$ layers. The case $4+4$, with $4$ localized spins in each layer, is identical to red solid line in Fig.~\ref{fig:fig3}(d) plotted here for easy comparison.}
		\label{fig:fig3size}
		\vspace{-0.5cm}
	\end{figure}

	The shrinking of the vector magnitude \mbox{$|\langle  \hat{\mathbf{S}}_i \rangle (t)|  < S\hbar$} is the consequence of entanglement of localized spins $i$ to all other degrees of freedom. Considering that in Fig.~\ref{fig:fig2} we study a closed total quantum system FM$_p$$\cup$FM$_a$$\cup$electrons that does not interact with natural intrinsic bosonic baths (such as phonons of the crystal lattice or photons of the ambient electromagnetic environment), external baths~\cite{Elbracht2020} or external fermionic reservoirs, this effect can arise solely from localized spin $i$ becoming entangled with other localized spins and/or  itinerant electron spin. To demonstrate this, in Fig.~\ref{fig:fig3} we examine the MI and the MLN between FM$_p$ and FM$_a$ layers, as defined by Eqs.~\eqref{eq:negativity}--\eqref{eq:entropy}. The increase of the MI in Fig.~\ref{fig:fig3}(c) and of the MLN in Fig.~\ref{fig:fig3}(d) from zero value at $t=0$ to finite value for $t>0$ confirms that the FM$_p$ and FM$_a$ layers are becoming dynamically entangled through nonequilibrium dynamics driven by quantum STT. The onset of such nonequilibrium and dynamical entanglement occurs at around \mbox{$t \simeq 20$ fs}, which is precisely the instant of time when the nonclassical dynamics of the localized spins of the FM$_a$ layer is initiated [bottom bundle of curves in Fig.~\ref{fig:fig2}(b)] or the von Neumann entropy of the FM$_a$ layer turns nonzero, $\mathcal{S}_{\mathrm{FM}_a}(t \gtrsim 20 \ \mathrm{fs}) > 0$, in Fig.~\ref{fig:fig3}(b).

	We also consider the localized spins of the FM$_p$$\cup$FM$_a$ subsystem at finite temperature.  In this case,  the von Neumann entropies and MI are nonzero in equilibrium (for $t \le 0$) even for separable (unentangled) mixed thermal state in Eq.~\eqref{eq:initialdmfinitet}, so we recompute only the MLN [dotted red line in Fig.~\ref{fig:fig3}(d)] at finite temperature \mbox{$T_\mathrm{Gibbs}=100$ K} since such quantity remains zero at $t=0$ [Fig.~\ref{fig:fig3}(d)]. The  thermal fluctuations of the localized spins are seen to postpone~\cite{Mondal2021} the onset of nonzero MLN to later times [dotted red line in Fig.~\ref{fig:fig3}(d)] when compared to MLN at zero temperature [solid black line in Fig.~\ref{fig:fig3}(d)]. They also lead to in smaller value of MLN when compared to the $T=0$ case, but without leading to its decay within simulated time.   
	
	In order to generate MLN decay and probe the underlying time scales, we couple each localized spin to bosonic bath at temperature $T_\mathrm{bb}$. Note that such 
	interactions, as exemplified by spin-phonon ones, are expected to be a major decoherence channel disrupting superpositions within mixed  entangled state $\hat{\rho}_{\mathrm{FM}_p \cup \mathrm{FM}_a}$. This produces the solid orange and red curves in Fig.~\ref{fig:fig3}(d), as computed [Sec.~\ref{sec:lindblad}] from  the Lindblad QME ~\cite{Lindblad1976,Manzano2020,Rudner2020,GarciaGaitan2023}. The decay of MLN, within the simulation time, is absent for \mbox{$T_\mathrm{bb}=10$ K}, while its presence for \mbox{$T_\mathrm{bb}=100$ K} suggests transient macroscopic entanglement on the scale of \mbox{$\sim 100$ fs}. This suggests  that conducting experiments on SVs at ultralow temperature, as already achieved in Ref.~\cite{Zholud2017} by using \mbox{$T \simeq 1$ K}, could generate entangled mixed states whose nonzero MLN lives for  sufficiently long time to enable detection by the scheme proposed in Sec.~\ref{sec:witnessing}. We note that Lindblad QME-computed curves in Fig.~\ref{fig:fig3}(d) exhibit an unphysical increase of MLN at early times, even before interaction with injected electron takes place, but this is simply an artifact~\cite{Daley2004} of quantum trajectories algorithm employed to solve the QME. In other words, entanglement between distant localized spins of two FM layers via dissipative environment, conjectured as possibility 
	in Ref.~\cite{Zou2022}, is not seen if we compute MLN at early times via computationall much more expensive  4$^\mathrm{th}$-order Runge-Kutta method.
	
	Finally, in Fig.~\ref{fig:fig3size} we examine the scaling of MLN with the total number of localized spins $N/2+N/2$ within FM$_p$$\cup$FM$_a$ subsystem of the SV in Fig.~\ref{fig:fig1}. The growth of entanglement with $N$ is in accord with time evolution of quantum many-body systems typically leading to a state with maximal entanglement (in the absence of localization by spatial disorder or nonunitary evolution due to projective measurements or decoherence by environment~\cite{Lu2021,Skinner2019}) allowed by  conservation laws [see  Sec.~\ref{sec:tomography}] and symmetries.
	
	\vspace{-0.4cm}
	\subsection{Quantum tomography of mixed entangled state of localized spins within SV}\label{sec:tomography}
	
\begin{figure}
		\includegraphics[width=\linewidth]{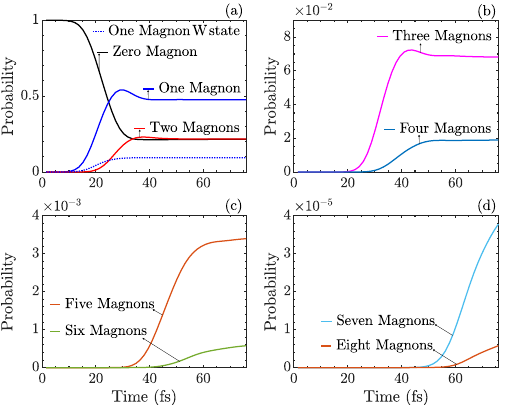}
		\caption{Time-dependence of probability [Eq.~\eqref{eq:prob}] to find localized spins at \mbox{$T=0$ K} of the FM$_p$$\cup$FM$_a$ subsystem within the SV in Fig.~\ref{fig:fig1} in the subspace spanned by many-body states $|\sigma_1 \ldots \sigma_{N/2} \rangle  |\sigma_{N/2+1} \ldots \sigma_N \rangle$ where $m=0, 1, 2, \ldots , 8$ spins are flipped with respect to the initial  state in Eq.~\eqref{eq:initialstate}. Such excited $m$-magnon  states are compatible with the spin conservation [Eq.~\eqref{eq:conserve}], and examples of projectors onto their subspaces are given in Eqs.~\eqref{eq:zeromag} and Eq.~\eqref{eq:onemag}. The time evolution is initiated by injecting a single-electron ($N_e=1$) current pulse, with other parameters the same as in Figs.~\ref{fig:fig2}--\ref{fig:fig3size}.}
		\label{fig:fig5}
		\vspace{-0.5cm}
	\end{figure}

	Quantum tomography is the task of reconstructing the full quantum state of a system from multiple measurements, which scales exponentially in the system size. Nevertheless, for sufficiently small many-body system it is possible to reconstruct, by measurement or from numerical calculations~\cite{Hemeniuk2019}, the structure of its density matrix. We delineate the structure of the density matrix as a function of time for mixed entangled state of all localized spins within the SV in Fig.~\ref{fig:fig1} using 4+4 case from Figs.~\ref{fig:fig2}---\ref{fig:fig3size}. For this purpose, specific sums of the diagonal elements of $\hat{\rho}_{\mathrm{FM}_p \cup \mathrm{FM}_a}(t)$  are plotted in Fig.~\ref{fig:fig5} [except for dotted line in Fig.~\ref{fig:fig5}(a) which involves off-diagonal elements, see last paragraph of this Section],   and both diagonal and off-diagonal elements are animated in the movie in the SM~\cite{sm}. In equilibrium $t \le 0$, the quantum state of the FM$_p$$\cup$FM$_a$ subsystem is given by  
	\vspace{-0.2cm}
	\begin{equation}\label{eq:initialstate}
		|\Sigma (t\!=\!0) \rangle_{\mathrm{FM}_p \cup \mathrm{FM}_a}\! = |\mathrm{FM}_p \rangle \otimes |\mathrm{FM}_a \rangle\! =\! |\!\! \uparrow \uparrow \uparrow \uparrow \rangle  |\!\! \downarrow \downarrow \downarrow \downarrow \rangle,
	\end{equation}
	being pure and separable (unentangled). The single-electron current pulse injection will initiate its evolution, necessarily converting it into mixed quantum state [Figs.~\ref{fig:fig3} and ~\ref{fig:fig3size}], as  described by the density matrix $\hat{\rho}_{\mathrm{FM}_p \cup \mathrm{FM}_a}(t)$. The diagonal matrix elements of  $\hat{\rho}_{\mathrm{FM}_p \cup \mathrm{FM}_a}(t)$ in proper representation give the probability to find the FM$_p$$\cup$FM$_a$ subsystem in a subspace of  quantum many-body state with zero  or $m$ spin-flips (i.e., magnons~\cite{Bajpai2021})
	\begin{equation}\label{eq:prob}
		\mathrm{Prob}_\mathrm{m-magnons}(t)=\mathrm{Tr}[\hat{P}_\mathrm{m-magnons} \hat{\rho}_{\mathrm{FM}_p \cup \mathrm{FM}_a}(t)],
	\end{equation}
	where the projection operators are given explicitly for 0- and 1-magnon cases as
	\begin{widetext}
		\vspace{-0.3cm}
		\begin{eqnarray}
			\hat{P}_\mathrm{0-magnon} & = & \ket{\up \up \up \up}\ket{\dn \dn \dn \dn}\bra{\up\up\up\up}\bra{\dn \dn \dn \dn}, \label{eq:zeromag} \\
			\hat{P}_\mathrm{1-magnon} & = &
			\ket{\df\up\up\up}\ket{\dn\dn\dn\dn}\bra{\df\up\up\up}\bra{\dn\dn\dn\dn} + 
			\ket{\up\df\up\up}\ket{\dn\dn\dn\dn}\bra{\up\df\up\up}\bra{\dn\dn\dn\dn} +
			\ket{\up\up\df\up}\ket{\dn\dn\dn\dn}\bra{\up\up\df\up}\bra{\dn\dn\dn\dn} \nonumber \\
			&+& \ket{\up\up\up\df}\ket{\dn\dn\dn\dn}\bra{\up\up\up\df}\bra{\dn\dn\dn\dn} + 
			\ket{\up\up\up\up}\ket{\uf\dn\dn\dn}\bra{\up\up\up\up}\bra{\uf\dn\dn\dn} +
			\ket{\up\up\up\up}\ket{\dn\uf\dn\dn}\bra{\up\up\up\up}\bra{\dn\uf\dn\dn} \nonumber \\ 
			&+&
			\ket{\up\up\up\up}\ket{\dn\dn\uf\dn}\bra{\up\up\up\up}\bra{\dn\dn\uf\dn} +
			\ket{\up\up\up\up}\ket{\dn\dn\dn\uf}\bra{\up\up\up\up}\bra{\dn\dn\dn\uf}. \label{eq:onemag}
		\end{eqnarray}.
		\vspace{-0.4cm}
	\end{widetext}

	Figure~\ref{fig:fig5} shows the time dependence  $\mathrm{Prob}_\mathrm{m-magnons}(t)$ [Eq.~\eqref{eq:prob}], for different $m=1,\ldots,8$ magnon states allowed by energy and angular momentum conservation laws~\cite{Mitrofanov2020,Mitrofanov2021}, in the course of the same time evolution studied in 
	Figs.~\ref{fig:fig2}--\ref{fig:fig3size}. That is, only those $m$-magnon which conserve total spin in the $z$-direction can be excited
	\begin{equation}\label{eq:conserve}
		[\hat{H},\hat{S}^z_\mathrm{tot}]=0,
	\end{equation}
	where $\hat{S}^z_\mathrm{tot}=\hat{s}_e^z + \hat{S}_1^z + \ldots + \hat{S}_N^z$ is the $z$-component of the total spin operator of electron and all localized spins. Note that $\langle \hat{S}^z_\mathrm{tot} \rangle (t)=0$ for the setup in Fig.~\ref{fig:fig1}. Figure~\ref{fig:fig5}(a)  shows that the initial state in Eq.~\eqref{eq:initialstate}, or equivalently $\mathrm{Prob}_\mathrm{0-magnons}(t)$, maintains its probability close to one, until $m=1,2,3$-magnon states are concurrently excited around \mbox{$t \simeq 20$ fs} in Figs.~\ref{fig:fig5}(a) and ~\ref{fig:fig5}(b). At later times, even state $\ket{\df\df\df\df}\ket{\uf\uf\uf\uf}$ with all eight spins flipped becomes excited, as signified by $\hat{P}_\mathrm{8-magnon} > 0$ in  Fig.~\ref{fig:fig5}(d). The probabilities of exciting $m \ge 3$ magnon states are progressively smaller in Figs.~\ref{fig:fig5}(b)--(d), when compared to $m=1,2$-magnon states in Fig.~\ref{fig:fig5}(a), but they would increase further $N_e>1$ electrons comprise the injected current pulse.  	The total number of excited $m$-magnon states compatible with conservation law in Eq.~\eqref{eq:conserve}  is $182$ out of $256$ for $4+4$ localized spins---for example, $\mathrm{Tr}\big(\ket{\df\df\df\df}\ket{\dn\dn\dn\dn}\bra{\df\df\df\df}\bra{\dn\dn\dn\dn} \hat{\rho}_{\mathrm{FM}_p \cup \mathrm{FM}_a}(t) \big) \equiv 0$ because such 4-magnon state  would violate Eq.~\eqref{eq:conserve}. Thus, in contrast to single FM layer driven by injection of fully spin-polarized electrons, where spin conservation [Eq.~\eqref{eq:conserve}] limits the number of different excitations  (such as 1-magnon for single injected electron~\cite{Mondal2019,Mitrofanov2021}), here $\langle \hat{S}^z_\mathrm{tot} \rangle (t) = 0$ allows for many joint excitations of FM$_p$ and FM$_a$ layers with initially antiparallel magnetizations, as long as their total spin-$z$ cancels (akin to many-spinon excitations by quantum STT on an antiferromagnet~\cite{Mitrofanov2021}).

	\begin{figure}[!t]
		\includegraphics[width=\linewidth]{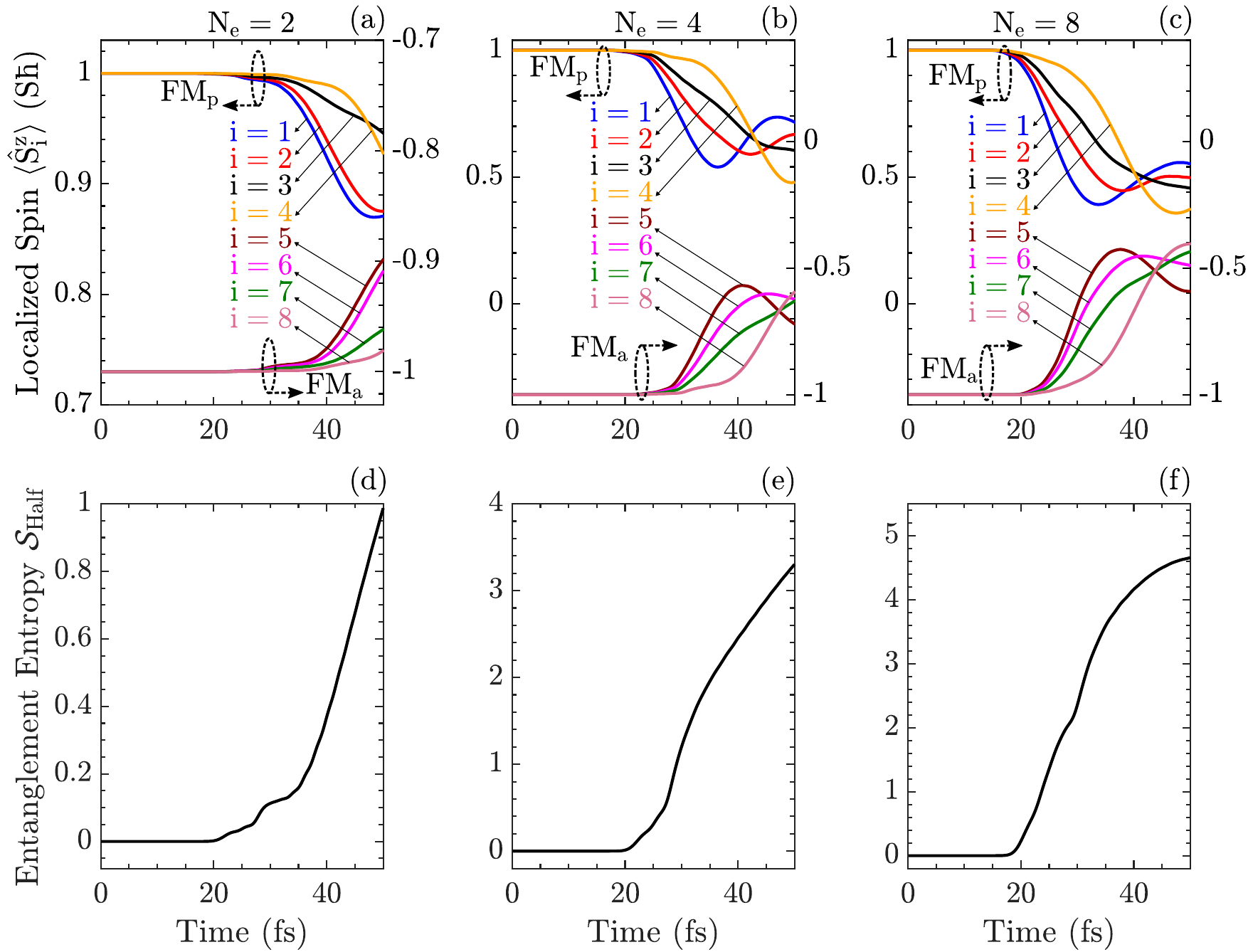}
		\vspace{-0.6cm}
		\caption{Time evolution (at \mbox{$T=0$ K}) using the tDMRG algorithm~\cite{White2004,Schmitteckert2004,Daley2004,Feiguin2011,Paeckel2019} of the expectation values of  localized spins initiated by injecting a spin unpolarized multi-electron ($N_e>1$) current pulse: (a) $N_e=2$; (b) $N_e=4$; and (c) $N_e=8$. The corresponding von Neumann entropy $\mathcal{S}_\mathrm{half}(t)$ [Eq.~\eqref{eq:entropy}] of the half of the system in Fig.~\ref{fig:fig1}, which includes all sites $i=57$--$112$ hosting localized spins of the FM$_a$ layer and all electrons present in that half at time $t$, is plotted in respective panels (d)--(f). The  initial quantum state of the whole system, with antiparallel magnetizations of the FM$_p$ and FM$_a$ layers of the SV in Fig.~\ref{fig:fig1}, is a pure many-body one with electrons confined within $N_\mathrm{conf}=10$ sites of the left edge of 1D chain in Fig.~\ref{fig:fig1} and with their total spin [Eq.~\eqref{eq:totalspin}] \mbox{$\langle \hat{\mathbf{s}}_e \rangle (t=0)$}.  For $t \ge 0$, the confining potential is switched off, so that electrons spread from left to right, thereby being injected into FM$_p$ layer. Localized spins $i=1$--$4$ belong to the FM$_p$ layer, and $i=5$--$8$ belong to the FM$_a$ layer.}
		\label{fig:fig4}
		\vspace{-0.7cm}
	\end{figure}

	By adding more localized spins and/or by injecting more electrons, larger and larger superpositions of separable states will be generated. They could be in principle written down explicitly for an arbitrary number of localized spins $N$, based on the conservation law in Eq.~\eqref{eq:conserve}, except that coefficients in front of individual terms in superposition comprising $\hat{\rho}_{\mathrm{FM}_p \cup \mathrm{FM}_a}$  have to be computed numerically, as depicted by Fig.~\ref{fig:fig5} and the movie in the SM~\cite{sm}. Such computation becomes impossible in the macroscopic limit $N \rightarrow \infty$. 
	We expect that such entangled states of macroscopically large number of degrees of freedom and with macroscopic number of terms in superpositions are more robust to decoherence, as it has been 
	studied~\cite{Aolita2015,Carvalho2004} in the context of entangled pure states as the number of localized 
	spin-$\frac{1}{2}$ (or, equivalently, qubits~\cite{Aolita2015,Carvalho2004}) increases $N \rightarrow \infty$. For example, these studies have concluded that the Greenberger-Horne-Zeilinger (GHZ) or ``cat'' state,  
	\mbox{$|\mathrm{GHZ}\rangle = \big(\! |\!\! \uparrow \uparrow \uparrow \uparrow \rangle  |\!\! \downarrow \downarrow \downarrow \downarrow \rangle + \! |\!\! \downarrow \downarrow \downarrow \downarrow \rangle  |\!\! \uparrow \uparrow \uparrow \uparrow \rangle \big)/\sqrt{2}$} written here for the case of our FM$_p$$\cup$FM$_a$,  exhibits much less robust entanglement than the so-called $W$ state, 
	$|W\rangle = \big( \ket{\df\up\up\up}\ket{\dn\dn\dn\dn} + \ket{\up\df\up\up}\ket{\dn\dn\dn\dn} + \ldots + \ket{\up\up\up\up}\ket{\dn\dn\dn\uf} \big)/\sqrt{8}$,
	with built-in size-robustness against several types of decoherence~\cite{Aolita2015,Carvalho2004}. Note that excitation of $W$ state is encoded by a $1$-magnon sector of the off-diagonal elements of $\hat{\rho}_{\mathrm{FM}_p \cup \mathrm{FM}_a}$, as confirmed by  nonzero probability  $\mathrm{Prob}_\mathrm{1-magnon-\mathrm{W}}(t)=\mathrm{Tr}\big( |W \!\rangle \langle W| \hat{\rho}_{\mathrm{FM}_p \cup \mathrm{FM}_a}(t) \big)$ plotted as dotted line in Fig.~\ref{fig:fig5}(a).   
	
	\vspace{-0.5cm}
	\subsection{Multi-electron current pulse injection}\label{sec:multi}
	
	The exact time evolution of the system in Fig.~\ref{fig:fig1} with more than one injected electron can, in principle, be obtained by brute force application of the time-evolution operator in Eq.~\eqref{eq:evolutionoperator}. However, such an approach is limited to very small systems due to the exponential growth of the basis with the system size. For example, for the SV modeled on $L_x=112$ sites (employed in tDMRG calculations in Fig.~\ref{fig:fig4}) and with $N=8$ localized spin-$\frac{1}{2}$ in the FM$_p$ and FM$_a$ layers, the vectors and matrices in Eq.~\eqref{eq:evolutionoperator} have size  $\binom{L_x}{N_e^\uparrow} \binom{L_x}{N_e^\downarrow} 2^{N}$. This reaches $\sim 10^{16}$ for $N_e^\uparrow=N_e^\downarrow=4$  and a total of $N_e= N_e^\uparrow+N_e^\downarrow=8$ electrons [employed in the pulse in Figs.~\ref{fig:fig4}(c) and ~\ref{fig:fig4}(f)].  To overcome this unfavorable scaling, we employ the adaptive tDMRG algorithm~\cite{White2004,Schmitteckert2004,Daley2004,Feiguin2011,Paeckel2019} for which the computational complexity is polynomial (instead of exponential) in system size.
	
	When compared to the single-electron ($N_e=1$) current pulse used in Fig.~\ref{fig:fig2}(b), the vector magnitude \mbox{$|\langle  \hat{\mathbf{S}}_i \rangle (t)|  < S\hbar$}, as a purely nonclassical effect driven by quantum STT~\cite{Petrovic2021a,Mondal2021}, shrinks further in Figs.~\ref{fig:fig4}(a)--(c) since more spin angular momentum can be transfered from the current pulse to the localized spins~\cite{Petrovic2021a}. This also leads to a progressively larger asymptotic value of the von Neumann entropy $\mathcal{S}_\mathrm{half}$ of the {\em half} of the lattice in Fig.~\ref{fig:fig1} of the total system  FM$_a$$\cup$FM$_p$$\cup$electrons, as the number of electrons comprising the pulse is increased from $N_e=2$ in Fig.~\ref{fig:fig4}(d) to $N_e=8$ in Fig.~\ref{fig:fig4}(f). Thus, the degree of macroscopic entanglement of two FM layers can be controlled by tailoring the magnitude and duration of injected electronic current pulse.

	\vspace{-0.3cm}
	\section{Witnessing nonequilibrium entanglement of localized spins via ultrafast X-ray spectroscopy}\label{sec:witnessing}
	
	Experimentally demonstrated protocols~\cite{Brydges2019,Bohnet2016} to quantify entanglement in quantum many-body systems use pure 
	quantum states of $\sim 10$ trapped cold atoms or ions, which can also be in a state of time-dependent nonequilibrium. Such protocols essentially 
	rely on unique capabilities of AMO techniques---such as the ability to make copies of a quantum system or single-atom manipulation and observation---which make it possible to isolate a subsystem (consisting of one or few atoms) and directly measure its (possibly time-dependent) R\'{e}nyi entanglement entropy~\cite{Brydges2019}.  However, they are limited to small system sizes and cannot be ported to solid state lattices that are always at finite temperature [hence, possible entangled states are necessarily mixed as in Eq.~\eqref{eq:mixedentangled}] and can also be out-of-equilibrium (as in the case of spintronic devices like the one in Fig.~\ref{fig:fig1}). How to measure entanglement of mixed quantum states is far less understood even in the context of AMO systems~\cite{Elben2020a}. 
	
	The very recent neutron scattering experiments~\cite{Mathew2020,Scheie2021,Laurell2021} have succeeded to directly witness intrinsic entanglement in solids~\cite{Brukner2006} present without any external driver. A salient example of such entanglement of  a macroscopically large number of localized spins in equilibrium is offered by quasi-1D quantum antiferromagnets~\cite{Mathew2020,Scheie2021,Laurell2021}. Witnessing of their entanglement was achieved by extracting the quantum Fisher information (QFI)~\cite{Toth2012,Hyllus2012} from neutron scattering data on  dynamic susceptibility~\cite{Hauke2016,Brukner2006}. However, this approach is not applicable to nanoscale devices like the SV in Fig.~\ref{fig:fig1} which are driven far from equilibrium by current pulses (dynamic susceptibility probes only near equilibrium states by relying on the Kubo linear-response theory~\cite{Hauke2016}). 

	\begin{figure}[!t]
		\includegraphics[width=\linewidth]{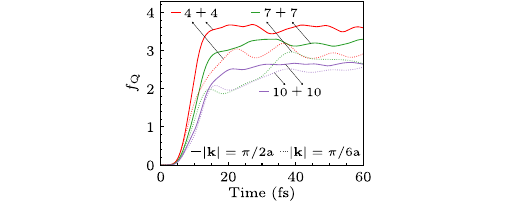}
		\caption{Time dependence of QFI $f_Q$ [Eq.~\eqref{eq:qfi}] for 4+4, 7+7 and 10+10 localized spins (at \mbox{$T=0$ K}) of the FM$_p$$\cup$FM$_a$ subsystem for momentum transfer $\hbar|\mathbf{k}|=\hbar\pi/6a$ (dotted lines) and $\hbar|\mathbf{k}|=\hbar\pi/2a$ (solid lines) between incident and scattered X-ray photons $\hbar \mathbf{k}=\hbar\mathbf{k}_i - \hbar\mathbf{k}_f$ in the setup of Fig.~\ref{fig:fig1}. The time evolution is initiated by injecting a single-electron ($N_e=1$) current pulse, with other parameters the same as in Figs.~\ref{fig:fig2}--\ref{fig:fig3size}.}
		\label{fig:fig6}
		\vspace{-0.55cm}
	\end{figure}

	Nonetheless, a very recently proposed scheme~\cite{Hales2022} makes it possible to obtain (via an integral equation) time-dependent QFI from transient dynamical structure factor $S(\mathbf{k},\omega,t)$. Although  $S(\mathbf{k},\omega,t)$ is directly proportional to inelastic neutron scattering cross section, neutron scattering at present offers slow time resolution \mbox{$\sim 1$ ms}. An ultrafast spectroscopy, with time resolution down to \mbox{$\sim 50$ fs} is offered by trRIXS~\cite{Mitrano2020,Chen2019,Halasz2016} as photon-in-photon-out (see Fig.~\ref{fig:fig1} for an illustration) scattering process with a resonant intermediate state and cross section that can be directly related~\cite{Hales2022} to $S(\mathbf{k},\omega,t)$. 
	
	The QFI is one of many entanglement witnesses~\cite{Scheie2021,Toth2012,Hyllus2012} which can distinguish  separable (unentangled) from entangled multipartite pure or mixed states. Although construction of a single witness that can detect all possible entangled states is computationally prohibitively expensive, QFI has been proven as particularly useful (e.g., it can witness multipartite entanglement even in complex topological quantum phases and their transitions~\cite{Pezze2017}) and robust against experimental artifacts~\cite{Scheie2021}.   
	The time-dependent QFI for the subsystem of localized spins of FM$_p$ and FM$_a$ layers in the SV in Fig.~\ref{fig:fig1} is defined by~\cite{Hales2022} 
	\begin{eqnarray}\label{eq:qfi}
		\lefteqn{f_Q(\mathbf{k},t)  =  \frac{4}{N} \sum_{i,j} e^{\mathbf{k} \cdot(\mathbf{r}_i - \mathbf{r}_j)}  \bigg\{ \mathrm{Tr} \big[\hat{S}_i^z \hat{S}_j^z  \hat{\rho}_{\mathrm{FM}_p \cup \mathrm{FM}_a}(t) \big]} \nonumber \\
		&& -  \mathrm{Tr} \big[\hat{S}_i^z  \hat{\rho}_{\mathrm{FM}_p \cup \mathrm{FM}_a}(t) \big] 
		\mathrm{Tr} \big[ \hat{S}_j^z  \hat{\rho}_{\mathrm{FM}_p \cup \mathrm{FM}_a}(t) \big] \bigg\},
	\end{eqnarray}
	where we replace instantaneous pure entangled quantum state used in Ref.~\cite{Hales2022} with our mixed entangled one  $\hat{\rho}_{\mathrm{FM}_p \cup \mathrm{FM}_a} (t)$. Here $\mathbf{r}_i$ is the real-space position vector of site $i$. The data from Figs.~\ref{fig:fig2}--\ref{fig:fig3size} can be reorganized via Eq.~\eqref{eq:qfi} to produce QFI for our SV  in Fig.~\ref{fig:fig1}, as shown in Fig.~\ref{fig:fig6}. We envisage that $\sim 1$~ns current pulse will ``pump'' the SV out of equilibrium via  quantum STT, so that delayed X-ray pulses applied during or after current pulse will be able to probe thereby induced nonequilibrium mixed entangled state $\hat{\rho}_{\mathrm{FM}_p \cup \mathrm{FM}_a}(t)$ with ultrafast temporal resolution during which decoherence cannot diminish MLN to zero [Fig.~\ref{fig:fig3}(d)]. Thus experimentally extracted 
	QFI can then be compared~\cite{Hales2022} with theoretical predictions~\cite{Halasz2016} exemplified by the one in Fig.~\ref{fig:fig6}.
	
	We note that such ``current-pump/X-ray-probe'' scheme is realistic. For example,  a similar one investigating the dynamics of localized spins driven by conventional STT has already been demonstrated in spintronics~\cite{Baumgartner2017} (using over $10^{12}$ current pulses of \mbox{$\sim 1$ ns} duration and time-resolved X-ray images with $100$~ps temporal resolution). However, in the probing of Ref.~\cite{Baumgartner2017} (based on X-ray magnetic circular dichroism rather than trRIXS) localized spins at room temperature are assumed to behave as classical vectors described by classical micromagnetics~\cite{Berkov2008} and the LLG equation,  which requires that their quantum state remains separable (unentangled) one, \mbox{$|\Sigma (t) \rangle_\mathrm{lspins} = |\sigma_1(t) \rangle \otimes |\sigma_2(t) \rangle \otimes \cdots |\sigma_N(t) \rangle$}, for all times $t$.

	\vspace{-0.3cm}
	\section{Conclusions and Outlook}\label{sec:conclusions}
	
	In conclusion, using the recently introduced concept of quantum STT~\cite{Zholud2017,Petrovic2021a,Mondal2019,Petrovic2021,Mitrofanov2020,Mitrofanov2021}, we predict that a spin-unpolarized current pulse comprised of flowing electrons can act as a mediator of macroscopic entanglement between two distant magnetic layers within a SV device. Thus generated entangled state is a mixed one [Figs.~\ref{fig:fig3} and \ref{fig:fig3size}], involving macroscopically large number of localized spins in realistic device. It would  also involve macroscopically large [Fig.~\ref{fig:fig5}] number of superimposed separable states because all such states that can be excited in accord with conservation laws operative in spin transfer are eventually introduced into the superposition over time. The latter feature can make them far more resistant~\cite{Aolita2015,Carvalho2004} to different types of decoherence than when only few separable states appear in superposition. 
	
	The SV is widely used device in fundamental and applied research in standard room-temperature spintronics, but for the entanglement scheme proposed here it will have to be kept at ultralow temperatures~\cite{Zholud2017}, \mbox{$T \lesssim 10$ K} according to Fig.~\ref{fig:fig3}(d),  in order to suppress decoherence and dissipation effects. This opens new avenue for exploration and manipulation of macroscopic entanglement where experimental demonstrations~\cite{Julsgaard2001,Ockeloen2018,Riedinger2018,Kotler2021,Lepinay2021,Lee2011,Thomas2021} thus far have been mostly focused on distant mechanical oscillators and photons as mediators. An advantageous nonequilibrium spintronic system for realizing our proposal is offered by ``intrinsic'' MTJs based on two-dimensional magnetic materials, such as two atomic planes of CrI$_3$ with oppositely oriented localized spins sandwiched by graphite electrodes~\cite{Song2018,Klein2018,Dolui2020}. This device naturally realizes the relative orientation of FM$_p$ and FM$_a$ layers in Fig.~\ref{fig:fig1}, while the spacer between them being vacuum and electrodes made of graphite  minimize the number of decoherence 
	chanels for the electronic orbital and spin degrees of freedom.

	We underline that the very recent neutron scattering experiments~\cite{Mathew2020,Scheie2021,Laurell2021} have demonstrated witnessing  entanglement of a macroscopically large number of localized spins within low-dimensional antiferromagnetic~\cite{Morimae2005,FrancisSong2011} materials in {\em equilibrium}, as long as they are kept below  ``entanglement temperature'' \mbox{$T_E \simeq 200$ K}~\cite{Scheie2021}. This achievement, previously demonstrated  only on few hundred trapped ions as simulators of quantum magnets~\cite{Brydges2019,Bohnet2016},  suggests that localized quantum spins in solids can be kept entangled even~\cite{Vedral2004,Morimae2005} when their number is macroscopically large and their temperature is finite~\cite{Mathew2020,Scheie2021,Laurell2021}.   Thus, we expect that our protocol for macroscopic entanglement could  work for {\em nonequilibrium} ferromagnets within SVs at sufficiently low  [Fig.~\ref{fig:fig3}(d)] temperatures  as well. While witnessing entanglement via QFI extracted from neutron scattering data for equilibrium bulk materials~\cite{Mathew2020,Scheie2021,Laurell2021} cannot be ported to nonequilibrium many-spin systems in our study, we envisage that 
	recently proposed~\cite{Hales2022} laser-pump/ultrafast-X-ray-probe scheme for bulk nonequilibrium materials extracting time-dependent QFI [Fig.~\ref{fig:fig6}] from trRIXS cross section can be adapted to spintronic devices as ``current-pump/X-ray-probe'' scheme. Such scheme is realistic, and some type of it has already been demonstrated in spintronics~\cite{Baumgartner2017} but for devices at room temperature with presumed classical dynamics of localized spins (as well as using different type of X-ray-probe). So, experimental challenge remains to develop ``current-pump/trRIXS-probe'' scheme for nonequilibrium spintronic device embedded into an external circuit and kept at ultralow temperatures.

	\begin{acknowledgments}
		We thank F. Garcia-Gaitan for illuminating discussions and help with the numerical implementation of Eqs.~\eqref{eq:ULE}--\eqref{eq:Spectral}. P.~M. and B.~K.~N. were supported by the U.S. National Science Foundation (NSF) Grant No. ECCS 1922689, and A.~S. was supported by the same foundation through the University of Delaware Materials Research Science and Engineering Center, Grant No. DMR-2011824. J.~P.~S.~P. and J.~M.~V.~P.~L. acknowledge financing by the Portuguese Foundation for Science and Technology (FCT) within the Strategic Funding UIDB/04650/2020 and through project No.\,POCI-01-0145-FEDER-028887. R.~D.~S and J.~P.~S.~P. are further funded by the Calouste Gulbenkian Foundation Foundation and FCT grant No.\,PD/BD/142774/2018, respectively. A.~E.~F. acknowledges the U.S. Department of Energy, Office of Basic Energy Sciences for support under grant No. DE-SC0014407.  A.~F. acknowledges support from the Royal Society (London) through a Royal Society University Research Fellowship. The supercomputing time was provided by the Viking Cluster (University of York) and DARWIN (Delaware Advanced Research Workforce and Innovation Network), which is supported by NSF Grant No. MRI-1919839.
	\end{acknowledgments}
	
	

\end{document}